\documentclass[11pt]{article}

\usepackage{graphicx}
\usepackage{amsmath}
\usepackage{dcolumn}
\usepackage{bm}

\begin{document}



\title{An algorithm for real-time estimation of Mezcal fermentation parameters based on redox potential measurements}

\author{P. Escalante-Minakata$^{\clubsuit}$,$\ $V. Ibarra-Junquera$^{\clubsuit}$,\\
H.C. Rosu$^{\spadesuit}$, A. De Le\'on-Rodr\'{\i}guez$^{\spadesuit}$ \& R. Gonz\'alez-Garc\'{\i}a$^{\diamondsuit}$ \\
\small $^{\clubsuit}$ Chemical Engineering Department, Faculty of Chemical Sciences,\\ \small University of Colima, Coquimatl\'an, Col., Mexico\\
\small $^{\spadesuit}$ Potosinian Institute of Science and Technology (IPICyT), San Luis Potos\'{\i}, S.L.P., Mexico\\
\small $^{\diamondsuit}$ State University of San Luis Potos\'{\i} (UASLP), San Luis Potos\'{\i}, S.L.P. Mexico\\
}

\maketitle

\begin{abstract}

\noindent {\small We describe an algorithm for the continuous
monitoring of the biomass and ethanol concentrations as well as the
growth rate in the Mezcal fermentation process. The algorithm
performs its task having available only the on-line measurements of
the redox potential. The procedure combines an artificial neural network
(ANN) that relates the redox potential to the ethanol and biomass
concentrations with a nonlinear-observer-based algorithm that uses the
ANN biomass estimations to infer the growth rate of this fermentation
process. The results show that the redox potential is a valuable
indicator of the metabolic activity of the microorganisms
during the Mezcal fermentation. In addition, the estimated growth rate can be
considered as a direct evidence of the presence of mixed culture
growth in the process. Usually, mixtures of microorganisms could be
intuitively clear in this kind of processes, however the total
biomass data do not provide definite evidence by themselves. In this
paper, the detailed design of the software sensor is presented, as
well as its experimental application at the laboratory level.}

\end{abstract}



\newpage

\section{Introduction}

Mezcal is a Mexican distilled spirit produced from the core of the
\textit{Agave} plant (the ``pi\~nas''). Several species are used for
the Mezcal production, i.e., \textit{Agave salmiana},
\textit{potatorum}, \textit{angustifolia}, \textit{tequilana}, etc,
and each produces a slightly different Mezcal. Therefore, the
tequila can be considered a regional type of Mezcal, restricted to
the usage of \textit{Agave tequilana} as raw material \cite{Tequila
Production}. It is important to mention that during this Mezcal
fermentative process the syrup obtained from the juice of cooked
``pi\~nas'' is left to naturally (spontaneously) ferment. By natural
alcoholic fermentations we refer to the ones that start by
themselves when a wide spectrum of different microorganisms starts
fermenting. Therefore, the fermentation of \textit{Agave} syrup into
Mezcal is a complex biochemical process involving a whole realm of
interactions between microorganisms.

During fermentation, the microorganisms employ sugars and other
constituents of \textit{Agave} syrup as substrate for their growth,
converting them into ethanol, carbon dioxide, higher alcohols and
their esters, and other metabolic compounds that contribute to the
chemical composition and sensory qualities of the Mezcal \cite{De
Leon-Rodriguez}. We also notice that the qualitative and
quantitative composition of the microbiota in fermenting musts could
depend on the following factors: region of the \textit{Agave}
origin, production procedure, initial cell concentration,
temperature, and ethanol concentration. As consequence, the
organoleptic properties are also the result of the diversity and
composition of microorganisms and their dynamics and frequency of
occurrence. Consequently, this fermentative process is a vital stage
in Mezcal making. Thus, it is very important to know more about the
dynamics of the entire microflora during the alcoholic fermentation
process. In other words, an algorithm that enables a monitoring
process could be fundamental for a quality control that ensures, at
least, homogeneity in the final product.

Nevertheless, a bottleneck in all biochemical monitoring process is
often the lack of sensors for biological variables. Moreover, it is
a well-known issue that in order to monitor many biotechnological
processes, the problem of growth rates estimations represent a
strategic feature. That is why several techniques have been
developed to estimate on line the biological variables from the
available measurements, which are usually dealing with
physicochemical variables. Depending on the obtainable information
about the process, there exist many possible types of estimators
that can be used \cite{Bastin 1986}, \cite{Locher 1992}, \cite{Farza
1998}.

Previously, various attempts of relating the \emph{redox potential}
to fermentation processes have been made taking into account that
redox potential assesses the growth ability of
microorganisms, as well as the physiological activity in a given
environments \cite{Kwong 1992}, \cite{Berovic 1999}, \cite{van Dijk
2000}, \cite{Cheraiti 2005}. Particularly, the practical
significance of redox potential and oxygen content at various stages
of winemaking was examined by \cite{Kukec 2002}. Many chemical,
enzymatic and biological processes in wine are correlated with the
oxidative state of the wine.

Monod kinetics, which were originally derived from laboratory
experiments with pure cultures and single substrates, are frequently
applied to describe the behavior of undefined mixed cultures growing
with single substrates or complex substrate mixtures \cite{180},
\cite{264}, \cite{83}. In this case, the growth parameters that have
been used represent overall values reflecting the growth constants
of many different strains with respect to the multicomponent
substrate and the frequencies and concentrations of both the
different substrates and microbial strains. Nevertheless, it is well
known from control processes that an accurate model leads to better
control design and therefore to better closed-loop performance.
There is recent progress shedding light on the dynamical processes
underlying the growth of mixed culture in a mixture of substrates
\cite{Reevesa 2004}, \cite{Ondeletas}. However, under such complex situations,
the accurate determination of the growth rates is still an open subject.

The work is organized as follows. Section \ref{Smat} is devoted to a
concise presentation of the fermentation experiment performed to
illustrate our approach. The software sensor, which is a combination
of an ANN and an adaptive observer scheme, is described in subsection
\ref{Ssoft}. The results obtained by applying this software sensor
to the Mezcal fermentation process together with a number of comments are
included in Section \ref{Sres}. The paper ends up with the
concluding remarks.

\section{Materials and Methods}\label{Smat}

\subsection{ Microorganism and culture conditions}

 In order to evaluate experimentally the performance of the estimation algorithm,
 we performed six individual batch experiments using inocula of
 native microorganisms (without the addition of any
commercial strain). The must (\emph{Agave} syrup) was
 obtained from \emph{Agave salmiana}, a species from the Mexican plateau
 (or altiplano) of the geographical region of San Luis Potos\'i.
 This must was centrifuged at  $8000 \ rpm \times 10 \ min$ and stored
 in a frozen state at $-20 ^{\circ}C$ prior to experiments.

 The batch fermentations were carried out in a bioreactor (Applikon, Schiedam, the Netherlands) of $1$
 liter. The bioreactor is equipped with pH and  redox sterilizable electrodes (Pt-Ring,
Applisens, Sensor Innovation, Applikon). The
 electrodes are connected to a console for data acquisition (Bioexpert, Data Acquisition Control Program, Version
 1.1x, Applicon), a device which is connected to a computer where the data
 are stored and computed.

 The bioreactor was filled with $900 \ mL$ of must as a culture medium,
 $100 \ mL$ of the inoculum in its exponential growing phase (biomass $0.1 \ g/L$)
  and $0.1$ \% of ammonium sulfate at final concentration.
The initial conditions of the fermentation were settled at a
temperature of $32.5 ^{\circ} C$ and initial sugar concentration of
$70 \ g/L$. The pH does not show a dynamic evolution, maintaining
itself at a value of $4$ during the whole process. For a schematic
representation of the experimental setup and the data acquisition system see
Fig.~\ref{FigBioreactor}.

\subsection{Analytical procedures and measurements}

 The batch processes have been monitored for $14$ hours, through sampling under sterile conditions. In order
to quantify biomass and ethanol concentrations, $5 \ mL$ samples
of culture was removed every $30\ min$. The samples were cleared
by centrifugation at $6000\ rpm$ for $5$ minutes at room
temperature. The next step was to collect the supernatant phase
and store it frozen at $-20 ^{\circ}C$ prior to be analyzed. The
obtained pellet was resuspended in distilled water in order to
proceed with the biomass analysis.

\subsubsection{Biomass and ethanol quantification}

The biomass measurements have been performed using (varian) UV
spectroscopy at $600 \ nm$. The obtained values were interpolated
with a standard curve of cell dry weight concentration.

For the determination of the ethanol, 1 $mL$ of each sample with a
final dilution 1:10 and 1 $mL$ of 1-butanol in vortex motion for
5 $min$ and followed by centrifugation at 6000 $rpm$ for 5
$min$. The organic phase was analyzed by gas chromatograpy (6890N
Agilent technologies, Wilmington, DE) provided with a capilar column
HP-Innowax ($30 \ m \times 0.25 \ mm$ i.d., $0.25 \ µm$ film
thickness; Agilent technologies, Wilmington, DE) and an auto-sampler
$7863$ (Agilent technologies, Wilmington, DE)  with a split relation
of $25:1$.

The chromatographic conditions were $35^{\circ}C$ for $2 \ min$,
increased at the rate of $10 ^{\circ}C/min$ up to $80 ^{\circ}C$,
and maintained at the latter temperature for $15 \ min$. The
carrier gas was helium at a flow rate of $1.5 \ mL/min$. The
temperatures of the injector and flame ionization detector (FID)
were set at $220 ^{\circ}C$ and $250 ^{\circ}C$, respectively. The
ethanol concentration of the samples was determined by means of a
calibration curve of known standard solutions of ethanol.

\begin{figure}[h]
\centering
          \includegraphics[height=8cm]{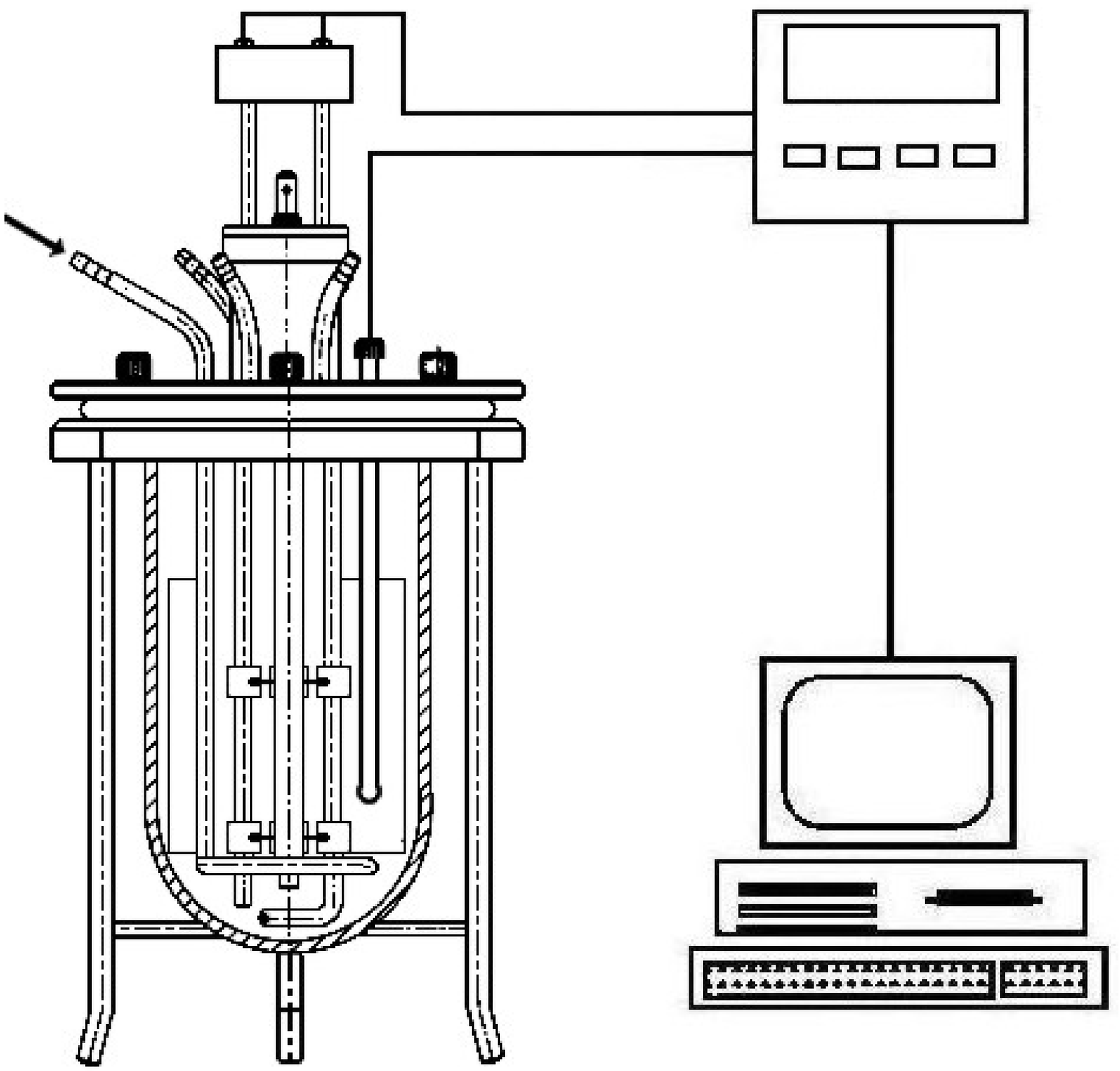}
            \put(-100,225){\footnotesize{Data acquisition device}}
            \put(-150,180){\footnotesize{\emph{redox potential}}}
            \put(-270,192){\footnotesize{\emph{Inlet stream of }}}
            \put(-280,180){\footnotesize{\emph{the heat exchanger}}}
            \put(-95,3){\footnotesize{Personal Computer}}
            \put(-108,-10){\footnotesize{with the Software Sensor}}
            \put(-195,-3){\footnotesize{Bioreactor}}
    \caption{The schematic representation of the experimental setup in our laboratory.}
    \label{FigBioreactor}
\end{figure}

\subsection{Redox potential}

The measurement of \emph{redox potential} is relatively fast,
accurate, and reliable and its values give an insight into the
oxidation process as well as the inherent ability of reduction in
the process, which is well established in the case of wine
\cite{Kukec 2002}. We mentioned that improvements in manufacturing
processes and equipment have allowed beer, wine, as well as Mezcal
producers to increase both the quantity and quality of these
products. It is crucial for the commercial competition to maintain
good quality control practices. Higher good-quality productions of
the manufacturing plants require faster and more sophisticated
analytical techniques to achieve them.

The measured values of the \emph{redox potential} can give
information on redox reactions in wine, which have an important
influence on its quality and stability \cite{Kukec 2002}. During
storing and aging of wine, oxidation and reduction processes affect
the character and taste of wine to a considerable extent \cite{Kukec
2002}, \cite{Cheraiti 2005}. We surmise that the same features could
occur in the case of the Mezcal making. Checking this assumption is
one of the main goals of this research. In the case of tequila,
there exist previous efforts to understand the relation of the
organoleptic properties with the process stages \cite{Benn},
\cite{Mercedez Capitulo}, \cite{Mercedez y Dufour},
\cite{Vallejo-Córdoba}. However, both for Tequila and Mezcal
there is no previous report of an online-measured variable that allows its
fermentation monitoring in real time.

The redox potential measurements were acquired periodically
each $0.01\ hr$ during $14 \ hrs$, and the data were stored and
computed on-line in a PC (see Fig.~\ref{FigBioreactor}).

\subsection{The Software Sensor}\label{Ssoft}

By the software sensor we mean the algorithm generated by the
coupling of the ANN and the adaptive observer. In this section we
develop step by step this type of algorithm. First, we present the
mathematical model that stays as the background of our approach.
Then we explain the relationship with the measured output, the
redox potential, and the ANN scheme. We end the section with
a discussion of the adaptive scheme used to infer the growth rate
function.

\subsubsection{Mathematical model of the fermentation process}

Batch microbial growth in a completely stirred bioreactor is
commonly described by the following differential equations
\cite{Bastin 1986}, \cite{Villadsen}
\begin{eqnarray}
\frac{ \mathrm{d} X_{1}}{ \mathrm{d} t}  &=& \  \mu (t) X_{1}  \nonumber\\
\frac{ \mathrm{d} X_{2}}{ \mathrm{d} t}  &=& \ -{k_1} \mu (t){X_{1}} \nonumber\\
\frac{ \mathrm{d} X_{3}}{ \mathrm{d} t}  &=& \ {k_2} \mu (t){X_{1}} \nonumber
\end{eqnarray}
where $X_{1}$ represent the total biomass concentration ($[g/l]$),
$X_{2}$ refers to the total substrate concentration ($[g/l]$) and
$X_{3}$ is the product concentration ($[g/l]$) (ethanol in this
case), ${k_1}$ represents the biomass yield, ${k_2}$ the product
yield and $\mu \left( t \right)$ is the growth rate. All the
parameters that have been used represent overall values reflecting
the many different strains with respect to the multicomponent
substrate. It is important to note that the scheme here proposed do
not need the values of ${k_1}$ or ${k_2}$ and $\mu (t)$
is inferred by means of the software sensor.

\subsubsection{Artificial neural network}

In order to relate the redox measurements to the ethanol and
biomass concentration an ANN procedure is applied. The methodology
that we carried out includes a forward-propagation training
algorithm for the ANN using some of our experimental data. In
order to perform our task we construct a model of the following
form:
\begin{eqnarray}
    X_{1}=f(X_{4})~,
\end{eqnarray}
\noindent where $X_{4}$ represents the redox potential
measurement data ($mV$), $X_{1}$ is the set of biomass concentration
data ($mg/L$) and the function $f(X_{4})$ is approximated by means
of the ANN procedure. The ANN architecture is of the standard type
\cite{lapedes} with a single ANN hidden-layer containing 10 units.
The same scheme was used for the case of the ethanol but using
$X_{3}=f(X_{4})$ as the ANN model.

Each unit of this network uses a sigmoid function as the activation
function. On the other hand, the output contains a linear activation
function, in our case the identity. The feed-forward training
algorithm considered here is the conjugate gradient method
\cite{rumel}. Three of the six individual batch experiments were
used to provide data for the training process. The error given by the ANN after the
training was as low as 0.0029.

\subsubsection{The adaptive observer}

The analysis of the adaptive observer scheme used here is based on
the following realistic assumptions:
\begin{itemize}
\item[(A1)] The specific growth rate $\mu (t)$  is
positive and bounded, that is an unknown $\mu_{max}$ exists
such that $0 < \mu (t) < \mu_{max}$.
\item[(A2)] There is no growth without substrate:
$  X_2 = 0  \Rightarrow  \mu (t) = 0$.
\item[(A3)] The time derivative of $\mu (t)$ is bounded:
$ |\frac{{\rm d} \mu (t)}{{\rm d} t}|  \leq M_1$, where $M_1 \in \mathrm{R}_{+}$.
\end{itemize}

\subsubsection{On-line estimation of $\mu(t)$ from measurements
of $X_1$}

Since $X_1$ is available through the neural algorithm mentioned
previously, we can now rewrite the output of the
system as:
\begin{eqnarray}
    y_s\,=\, X_1,
\end{eqnarray}
where $y_s$ is the set of on-line measurements of the system which
is available indirectly through ANN means. Then, according to the discussion in
\cite{Bastin 1986}, the following algorithm can be used to
estimate $\mu(t)$:
\begin{eqnarray}
\frac{ \mathrm{d}\ \hat{X}_{1}}{ \mathrm{d} t}  &=& \  y_s \
\hat{\mu} \left( t \right) +
\mathcal{K}_1\,y_s \left ( y_s- \hat{X}_1 \right)\nonumber\nonumber\\
\frac{ \mathrm{d}\ \hat{\mu}}{\mathrm{d} t}\  &=& \
\mathcal{K}_2\,y_s \left ( y_s- \hat{X}_1 \right),\nonumber
\end{eqnarray}
where $\hat{X}_1$ and $\hat{\mu}$ represent the estimated value of
$X_1$ and $\mu$, respectively.
The constants
$\mathcal{K}_1$ and $\mathcal{K}_2$ must be chosen such that:
\begin{eqnarray}
    0\,<\,\mathcal{K}_2\,<\,\frac{\mathcal{K}_1^2}{4}.
\end{eqnarray}
The above condition ensures the asymptotic convergence of the
observer error to a neighborhood of zero (\cite{Bastin 1986}).

\section{Results and Discussion}\label{Sres}

In the previous section, the general idea as well as the detailed
steps for the construction of the software sensor have been given.
To complete the analysis of the approach, this section presents
the experimental results obtained at the laboratory level at which
the software sensor was tested.

\begin{figure}[h]
      \begin{minipage}[t]{.5\textwidth}
    \begin{center}
    \includegraphics[height=5cm]{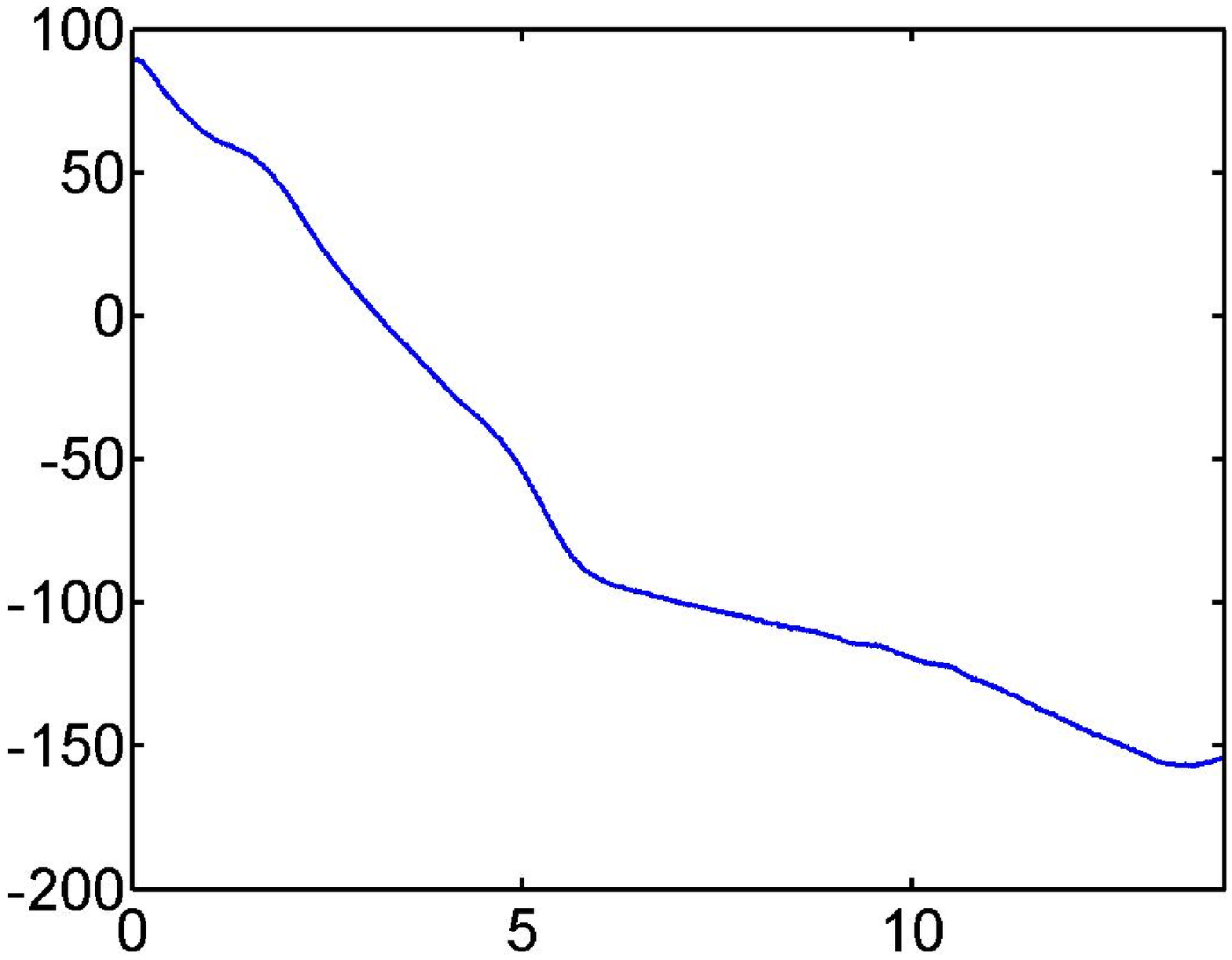}
        \put(-193,20){\rotatebox{90}{\small{Redox Potential (mV)}}}
        \put(-100,-5){\small{time ($hr$)}}
 \end{center}
  \end{minipage}
  \hfill
  \begin{minipage}[t]{.5\textwidth}
    \begin{center}
 \includegraphics[height=5cm]{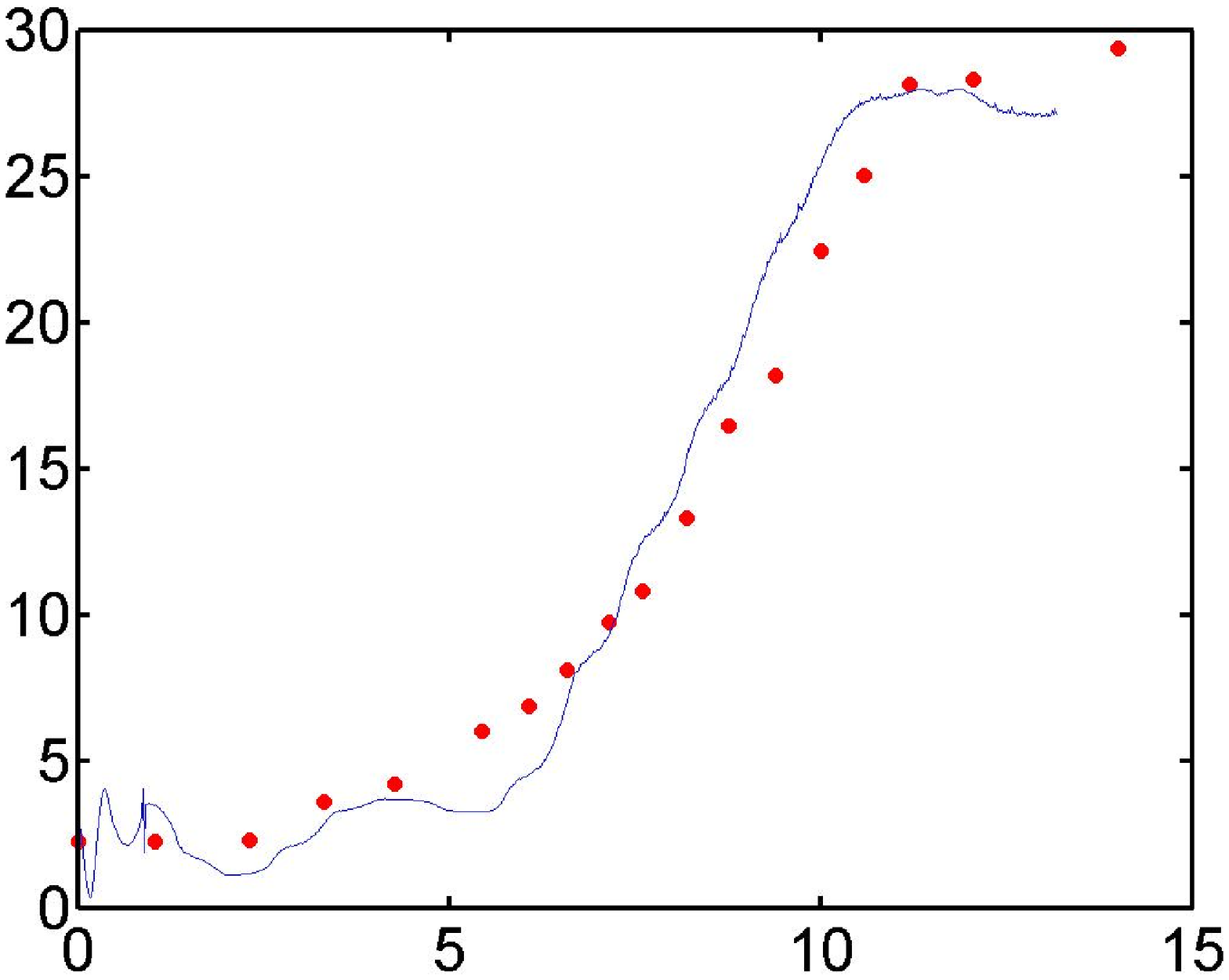}
        \put(-193,25){\rotatebox{90}{\small{Ethanol Conc. $(g/L)$}}}
        \put(-100,-5){\small{time ($hr$)}}
    \end{center}
  \end{minipage}
  \hfill
    \caption{The right plot illustrates the time evolution of the redox potential
    in the Mezcal fermentation process. In the left plot the blue solid line represents the
    ANN-estimated ethanol concentration while the red dots stand for the experimental values.}
    \label{figANN}
\end{figure}

\begin{figure}[h]
\centering
    \includegraphics[height=7cm]{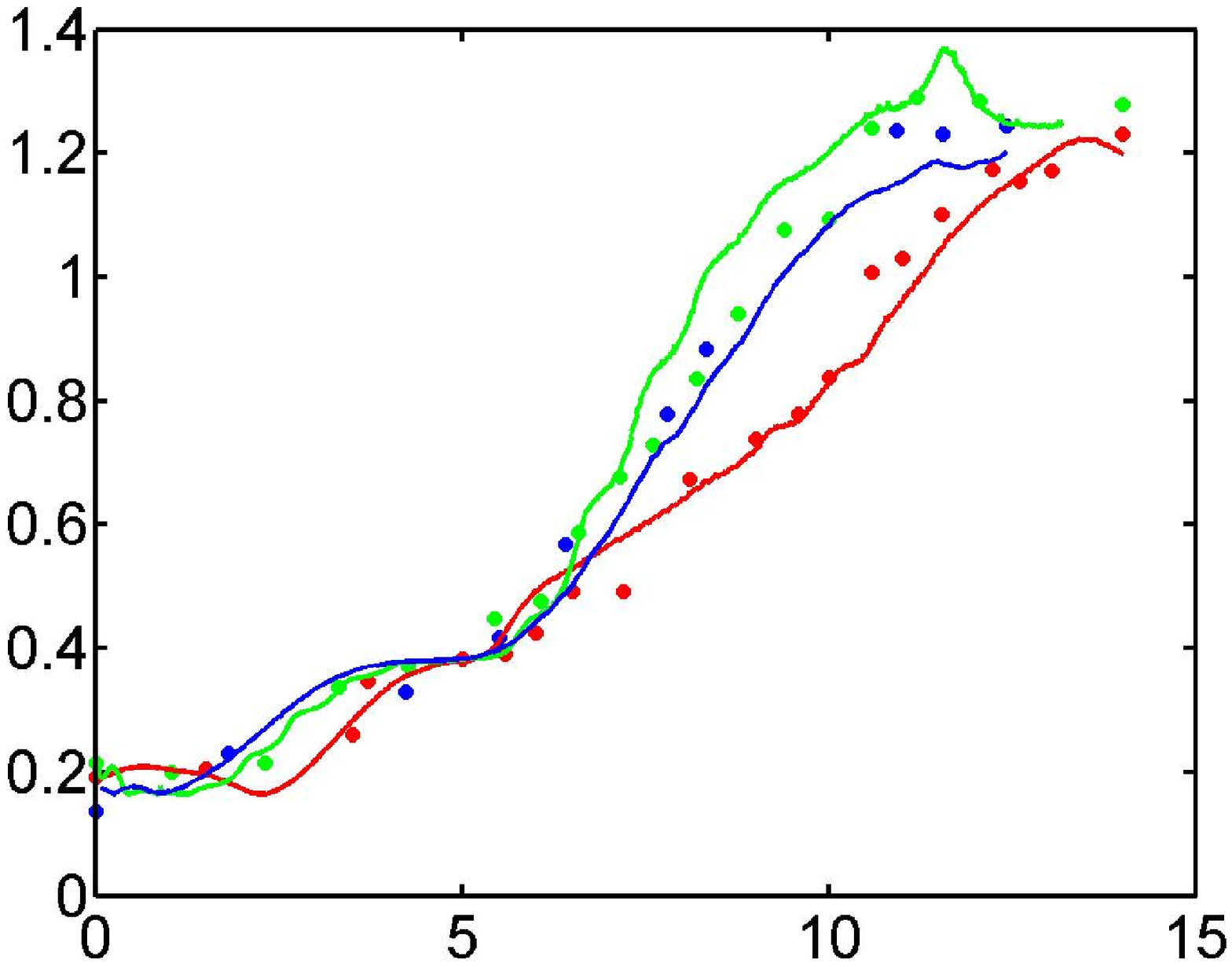}
        \put(-260,55){\rotatebox{90}{\small{Biomass Conc. $(g/L)$}}}
        \put(-150,-5){\small{time  ($hr$)}}
    \caption{The continuous lines correspond to the estimated biomass
    data and the dots to the experimental data. Blue, red and green colors correspond to
three of the experiments carried out in our laboratory.}
    \label{FigEstiamtedBiomass}
\end{figure}

In total, six series of experiments of Mezcal fermentation at
temperature of $32 ^{\circ}C$ were performed. The temperature has
been maintained constant using a heat exchanger device, see Fig.
\ref{FigBioreactor}. First, we performed three experiments to generate
the necessary biomass and ethanol concentration data for the
training of the ANN. Once the error given by the ANN goes below the
value of $0.003$ biomass units, the adaptive scheme was added to
complete the software sensor procedure. Then, three more experiments
were carried out for further testing the scheme. The results
concerned with the performance of the ANN to infer the ethanol and
biomass concentrations from the redox potential measurements
are presented in Figs.~\ref{figANN} and \ref{FigEstiamtedBiomass}, where
one can appreciate the degree of accuracy of the algorithm. Notice
that Fig. \ref{FigEstiamtedBiomass} presents several slope
changes that do not allow a direct inference of the presence of a mixture of
microorganisms during the fermentation process.

\begin{figure}[h]
\centering
    \includegraphics[height=3.1cm]{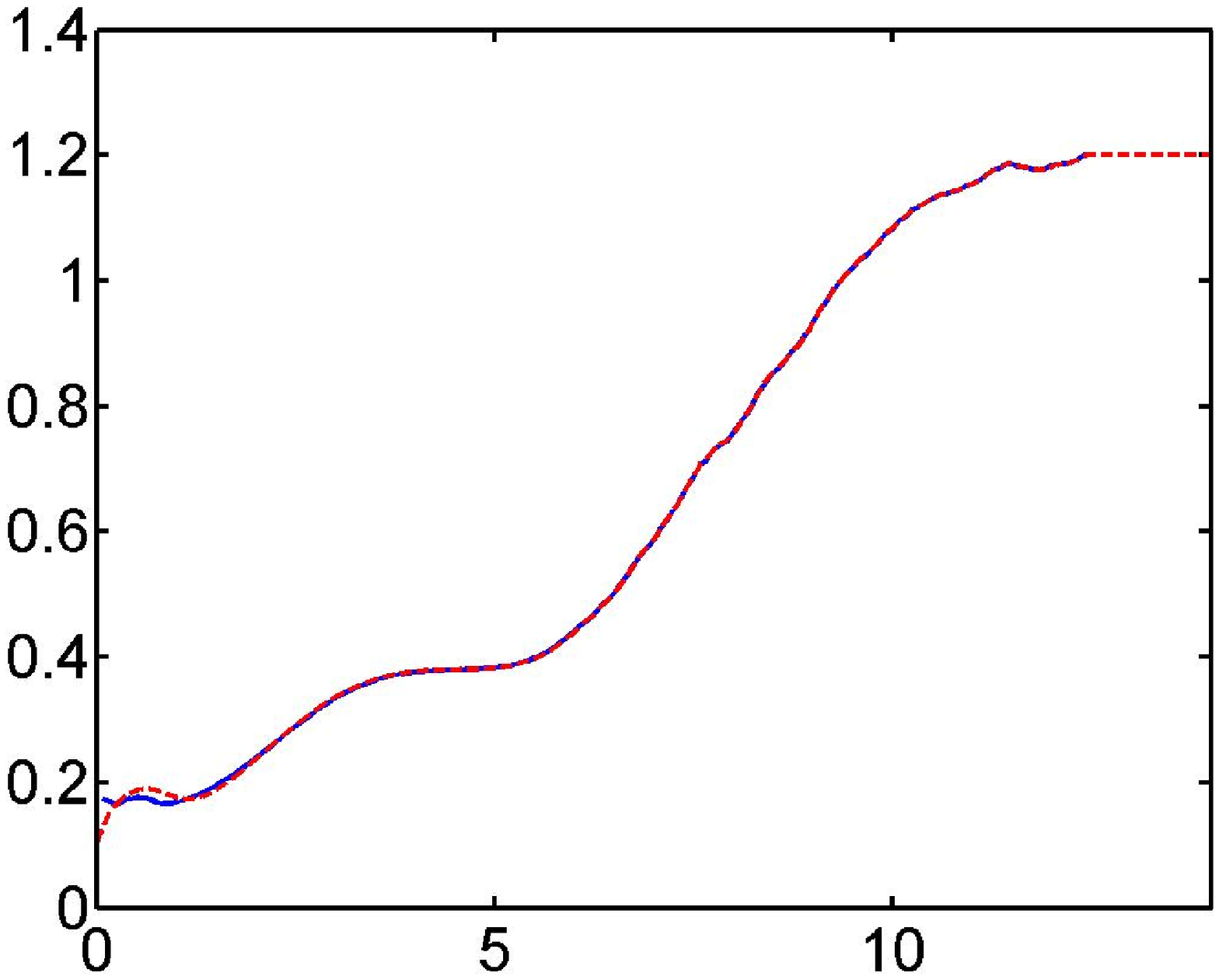}
    \includegraphics[height=3.1cm]{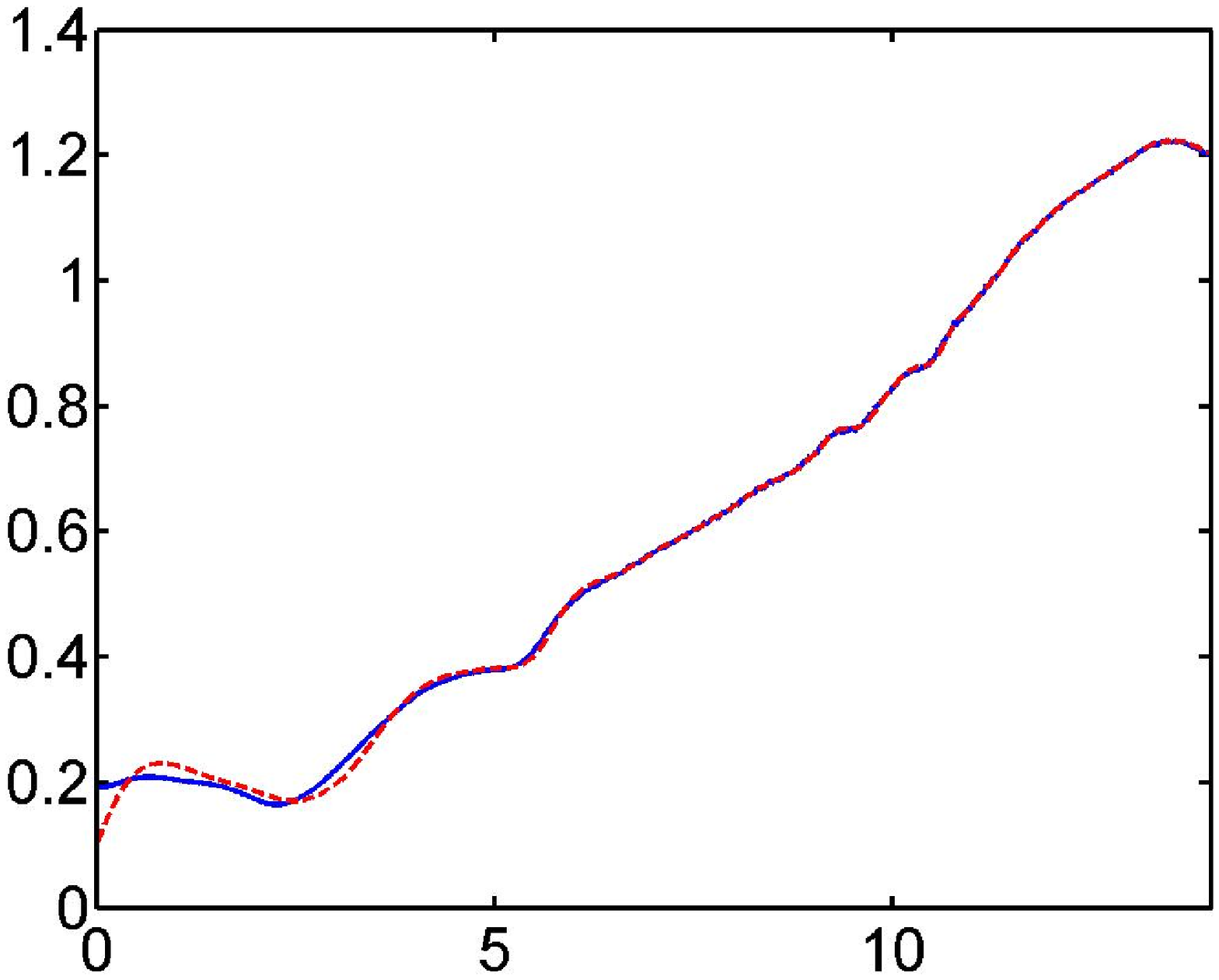}
    \includegraphics[height=3.1cm]{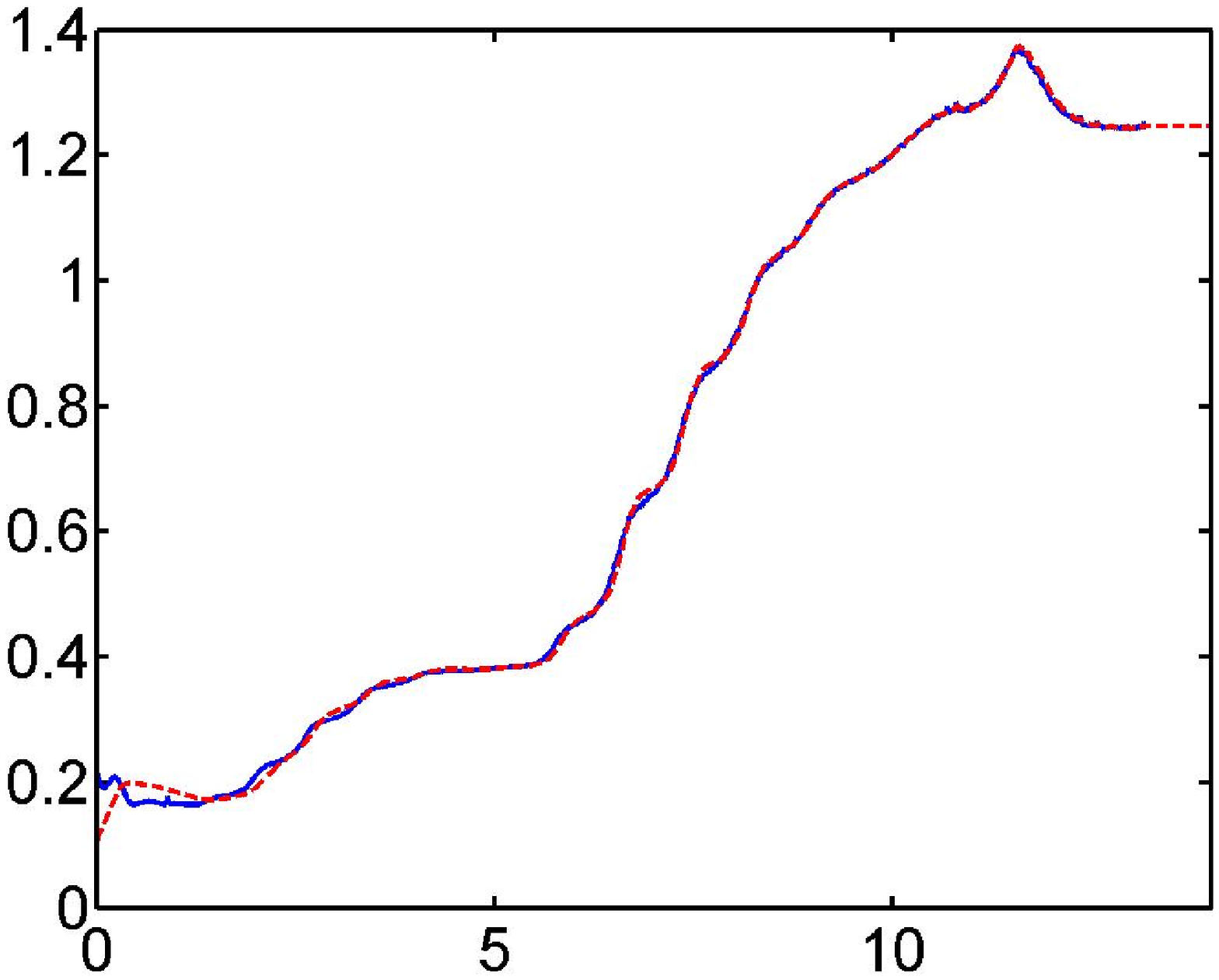}
     \put(-200,-5){\small{time  ($hr$)}}
     \put(-393,15){\rotatebox{90}{\tiny{Biomass Conc. $(g/L)$}}}
     \caption{(Blue) solid lines represent the estimated biomass concentration given by the software sensor and (red) dash-dotted lines
   stand for the ANN-predicted biomass concentrations. From left to right, the displayed plots correspond to the three experiment performed to test our approach.}
    \label{FigBiomassBySoftware}
\end{figure}

Although the main goal is to estimate $\mu (t)$, the biomass
estimation is also performed, and the difference between this
estimation and the value predicted by the ANN procedure is used as
a correction term in the algorithm. Figure~\ref{FigBiomassBySoftware}
shows the ability of the software
sensor to rebuild the biomass concentration data. It should be
highlighted that the simplicity associated with the implementation
of this algorithm and the fact that a single measured signal,
the redox potential, is needed are very promising features from the
technological and industrial standpoints.

\begin{figure}[h]
\centering
          \includegraphics[height=8cm]{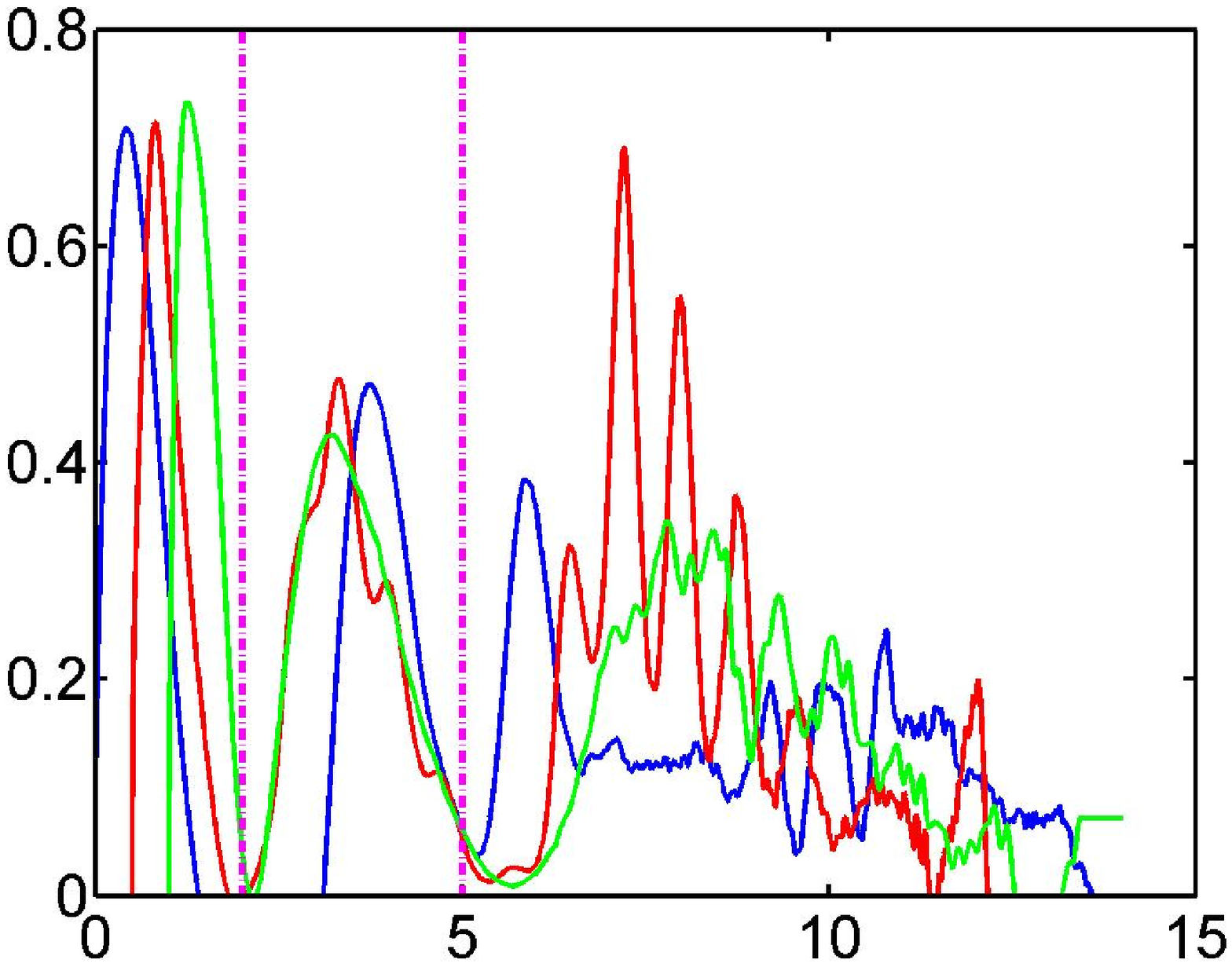}
        \put(-300,80){\rotatebox{90}{{growth rates ($\mu(t)$)}}}
        \put(-200,-5){\small{time ($hr$)}}
        \put(-252,198){$\mathcal{A}$}
        \put(-217,198){$\mathcal{B}$}
        \put(-120,198){$\mathcal{C}$}
    \caption{The growth rates as estimated by the adaptive algorithm. The blue, red an green lines correspond to the estimated results for
each of the experiments carried out in the bioreactor.}
    \label{mu estimated}
\end{figure}

In Fig.~\ref{mu estimated}, it is possible to distinguish three
regions, labeled as $\mathcal{A}$, $\mathcal{B}$ and $\mathcal{C}$.
In general terms, in region $\mathcal{A}$ one is not able to get any
conclusion on the dynamical behavior of the process since during the corresponding
time span the observer has a transient behavior. The span of time corresponding to region $\mathcal{A}$ is
given by the time that the observer takes up to minimize the error
between its estimated biomass concentration and that given by the ANN
procedure.

On the other hand, we associate regions $\mathcal{B}$ and
$\mathcal{C}$ with the presence of two different groups of
microorganisms, namely bacteria and yeasts. This assumption is
based on the fact that the growth rate of bacteria is faster than
that of yeasts. Thus, from Fig.~\ref{mu estimated}, the presence
of a mixed culture growth comes out naturally, a fact which is not
so obvious when one examines only the biomass data given in
Fig.~\ref{FigEstiamtedBiomass}. Note in addition that the end of
the fermentation process is quite clear in Fig.~\ref{mu
estimated}.

\section{Conclusions}\label{Sconc}

In this work, we have shown that in the Mezcal fermentation process
the redox potential can give relevant information on the
metabolic activity of the involved microorganisms, including both ethanol and biomass
concentration. Besides, the problem of estimating the specific
growth rates in the Mezcal fermentation process has been addressed through a
combination of a simple ANN procedure and a nonlinear observer algorithm.
Moreover, the strategy presented here detects in a clear way the end of
this fermentative process. The latter fact is quite relevant from
the production point of view since it is the basic piece of information by
which one can save time and avoid the degrading process due to the
conversion of the ethanol into acetic acid. This leads to an improvement in the
quality of the product. In addition, the computational scheme provides
a very appropriate tool for a fast and reliable quality control and can be used
to ensure the homogeneity of the final product. The methodology presented in this
paper is general and can be also used in automatic control applications.

\end{document}